%% file: paper.tex
\documentclass[conference, compsoc]{IEEEtran}
\IEEEoverridecommandlockouts
\usepackage{subfig}
\usepackage{caption}

\usepackage{amsmath}
\usepackage{booktabs}
\usepackage{graphicx}
\usepackage{multirow}
\usepackage{courier}
\usepackage{multirow}
\usepackage{acronym}
\usepackage{color}

\acrodef{IoT}{Internet of Things}

\begin{document}

\title{The Quest for Scalability and Accuracy:\\ Multi-Level Simulation of the Internet of Things\footnotemark}

\author{\IEEEauthorblockN{Stefano Ferretti, Gabriele D'Angelo, Vittorio Ghini, Moreno Marzolla}
\IEEEauthorblockA{Department of Computer Science and Engineering (DISI), University of Bologna, Italy\\
\{s.ferretti, g.dangelo, vittorio.ghini, moreno.marzolla\}@unibo.it}
}


\maketitle

\footnotetext{The publisher version of this paper is available at \url{https://doi.org/10.1109/DISTRA.2017.8167672}.
\textbf{{\color{red}Please cite this paper as: ``Stefano Ferretti, Gabriele D'Angelo, Vittorio Ghini, Moreno Marzolla. The Quest for Scalability and Accuracy in the Simulation of the Internet of Things: an Approach based on Multi-Level Simulation. Proceedings of the IEEE/ACM International Symposium on Distributed Simulation and Real Time Applications (DS-RT 2017)''.}}}

\begin{abstract}
This paper presents a methodology for simulating the Internet of
Things~(IoT) using multi-level simulation models. With respect to
conventional simulators, this approach allows us to tune the level of
detail of different parts of the model without compromising the
scalability of the simulation. As a use case, we have developed a
two-level simulator to study the deployment of smart services over
rural territories. The higher level is base on a coarse grained,
agent-based adaptive parallel and distributed simulator. When needed,
this simulator spawns OMNeT++ model instances to evaluate in more
detail the issues concerned with wireless communications in restricted
areas of the simulated world. The performance evaluation confirms the
viability of multi-level simulations for IoT environments.
\end{abstract}


\input{sec_introduction.tex}
\input{sec_related.tex}
\input{sec_multilevel.tex}
\input{sec_simulator.tex}
\input{sec_perf.tex}
\input{sec_conclusions.tex}

\small{
\bibliographystyle{abbrv}
\bibliography{paper}  
}

\end{document}

%% file: sec_introduction.tex
\section{Introduction}\label{sec:intro}

The use of the~\ac{IoT} to create and deploy smart services at a large
scale requires novel mechanisms to model things and understand how
these can interact to create effective solutions for data retrieval,
dissemination and
computation~\cite{iot_survey,Atzori:2010,gda-simpat-2014}. A multitude of
mobile terminals, sensors, RFID and other devices is being designed to
this aim~\cite{singh2014survey}. The heterogeneity among the devices
and the number of possible interaction scenarios makes it difficult to
understand all aspects concerned with any large instance of
the~\ac{IoT}. Thus, simulation can be very useful to get a better
understanding about quantitative and qualitative aspects of the system
under investigation. Simulation results can be used for capacity
planning, what-if simulation and analysis, proactive management and to
support security-related evaluations, just to name a few. However, the
huge scale of the~\ac{IoT} represents a major challenge to the usage
of existing simulation tools. Indeed, the conventional serial
approaches are often unable to scale to the number of nodes and level
of detail required by a large system such as the~\ac{IoT}.

To address this challenge we propose a multi-level simulation approach
for modeling large~\ac{IoT} environments that does not impose
over-simplification of the
model~\cite{gda-hpcs-17,magne2000towards}. Specifically, we use a ``high level''
simulator that provides a coarse-grained level of detail; this
simulator coordinates the execution of a set of domain specific
``lower level'' simulators that are spawned and executed where and
when a finer-grained level of detail is needed. The simulators at both
levels interact and synchronize their execution to compute correct
state updates. The low level simulators continue their job until
requested by the high level one. Then, the higher level simulator
updates the state, based on the outcomes from the lower level and
continues its computation. This means that the detailed simulation is
performed only on a portion of the simulated area and only when
needed.

As a use case, we study the development of smart services to be
deployed over decentralized territories, which are not provided with a
full wireless cellular network coverage~\cite{smartshires}. In
particular, we describe the implementation of a multi-level approach
to simulate a large number of (static or mobile) entities that
generate data that must be disseminated over the network using
wireless broadcast protocols. We assume that the data being
disseminated are enriched with tags, keys or metadata that allow nodes
to filter such data. In this way it is possible to design viable
publish-subscribe schemes that represent the substrate upon which
proximity-based applications can be developed. For example, a user
might decide to move toward a certain geographical area based on the
received information, e.g., to reach a certain producer offering some
peculiar product.


The above use case allows us to assess the ability of the proposed
multi-level simulation approach to scale to large, highly intensive
interacting~\ac{IoT} scenarios. Experimental evaluation shows that
multi-level simulation provides better scalability with respect to a
classic fine grained simulation, while offering the same level of
detail when needed. Therefore, multi-level simulation is a viable tool
to simulate IoT-based applications at a large scale.

The remainder of this paper is organized as follows.
Section~\ref{sec:related} presents some related
work. Section~\ref{sec:multilevel} briefly introduces issues concerned
with multi-level simulation.  In Section~\ref{sec:simulator} we
provide a discussion on the approach we propose to simulate large
scale~\ac{IoT} environments using multi-level simulation.  In
Section~\ref{sec:perf} a performance evaluation is shown.  Finally,
concluding remarks are discussed in Section~\ref{sec:conclusions}.

%% file: sec_related.tex
\section{Related Work}\label{sec:related}

It has been observed~\cite{gda-hpcs-16} that the simulation of
the~\ac{IoT} can generate a huge amount of data, depending on the
required level of detail. Low-level, detailed network simulators such
as OMNeT++, ns-2 or ns-3, when used alone are quite often inadequate
for that purpose~\cite{6069710} due to their lack of scalability.

Conversely, agent-based simulators (e.g., MASON~\cite{Luke:2005},
SUMO~\cite{SUMO2012}) are the preferred choice when one needs to
create models that mimic wide area (e.g.~urban) systems in
general. Agent-based simulators allow the user to select the time and
spatial scales of the model~\cite{Karnouskos}; therefore, these tools
have been used to study intelligent traffic control systems, mobile
applications and other similar
systems~\cite{bauza,kerekes,Wegener:2008,e16052384}. While agent-based
models can typically scale with respect to the size of the system
(e.g., number of network nodes, surface of the geographical area),
they do not allow to easily select the level of details of different
parts of the simulation, forcing the user to over-simplify the
model. This is not always a good solution, especially when dealing
with~\ac{IoT}, where networking issues may represent important local
aspects to consider.

It is also worth noticing that discrete event simulation approaches
are the common choice to simulate~\ac{IoT}
\cite{Brambilla:2014:SPL:2694768.2694780,6803226,6844677,paper+kirsche-13:iot-simulation}. However,
other kinds of simulation have been utilized in the literature.  For
example, MAMMotH is a software architecture based on emulation
\cite{6664581}, while Monte Carlo methods are employed
in~\cite{6418824}. Another proposed solution resorts to a model-driven
simulation, according to which the standard~SDL language is used to
describe an~\ac{IoT} scenario~\cite{Brumbulli2016}. Then, an automatic
code generation transforms the description into an executable
simulation model for ns-3.

Going back to discrete event simulation approaches,
in~\cite{Brambilla:2014:SPL:2694768.2694780} the authors propose to
integrate the~DEUS general-purpose discrete event simulation with the
domain specific simulators~Cooja and ns-3 for implementing
large-scale~\ac{IoT} simulations. This solution allows offering a good
scalability.

DPWSim uses models described through the OASIS standard ``Devices
Profile for Web Services'' (DPWS)~\cite{6803226}. Its main goal is to
provide a cross-platform and easy-to-use assessment of DPWS devices
and protocols. Unfortunately, it is not designed for very large-scale
setups.

The use of Cloud computing solutions enables large scale simulation
settings. In this sense, our solution would benefit from the
deployment of our simulator in a Cloud environment. As concerns other
proposals, SimIoT exploits Cloud environments for back-end
operations~\cite{6844677}; authors focus on a health monitoring system
for emergency situations in which short range and wireless
communication devices are used to monitor the health of patients.

Finally, in~\cite{paper+kirsche-13:iot-simulation} a hybrid simulation
environment is used where Cooja-based simulations (i.e., system level)
are integrated with a domain specific network simulator (i.e.,
OMNeT++).  In some sense, this approach resembles the idea of
exploiting a multi-level simulation, since it tries to keep the best
of the considered simulation approaches.

To sum up, it is evident that running a whole model representing
a~\ac{IoT} scenario at the highest level of detail is unfeasible.  A
better approach is to bind different simulators together, each one
running at its appropriate level of detail and with specific
characteristics to be simulated (e.g., mobility models, wireless/wired
communications). Our proposed approach, described in the next
sections, follows this idea.

%% file: sec_multilevel.tex
\section{Multi-level Simulation Architecture}\label{sec:multilevel}

The rationale behind multi-level simulation is to take multiple models
and glue them together into a hierarchical
structure~\cite{gda-hpcs-16,magne2000towards,gda-simpat-iot}, where
each simulator works at a different level of detail. A ``high level''
simulator starts the simulation and works at a coarse grained level of
detail. When and where a more detailed model is needed, the high level
simulator activates and coordinates the execution of one or more lower
level simulators. Such coordination capabilities might be performed by
the higher level simulator (this is the approach we will employ), or
by some external coordinator module.

\begin{figure*}[ht]
\centering
\includegraphics[width=.7\linewidth]{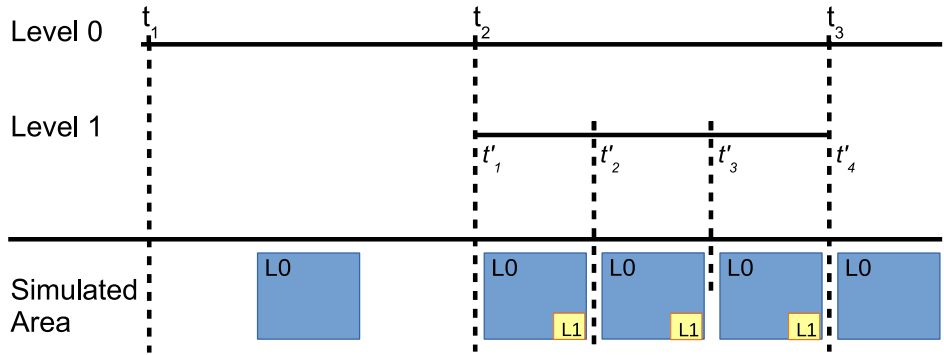}
\caption{Simplified multilevel simulation.}
\label{fig:multilevel}
\end{figure*}

Figure~\ref{fig:multilevel} shows an example of a two-level
simulation. Initially (timestep~$t_1$ in the figure) the model is
executed at a coarse level of detail by the top level (Level~0)
simulator. However, at a certain point in time ($t_2$), part of the
simulated scenario requires a finer level of detail. To achieve this,
a Level~1 simulator is activated to handle that part of the model and
all entities therein in more details -- for example, using a shorter
time step -- while everything else is still managed by the Level~0
simulator.  All the entities managed at Level~0 evolve using $t$-sized
(coarse grained) timesteps and all the others use $t'$-sized (fine
grained) timesteps. Timestep $t_2$ (that is the same of $t'_1$ for
Level 1) is the moment in which a part of the model components is
transferred from the coarse grained simulator to the finer one.  Then,
the components at Level 0 will jump from $t_2$ to $t_3$ while the
components simulated at Level 1 will be updated at $t'_2$, $t'_3$ and
$t'_4$.  When there is no more need for a finer level of detail, all
the components simulated at Level 1 are transferred again to the Level
0 simulator.

This approach clearly requires the ability for the simulators to
interoperate and synchronize. In the rest of the paper, we show how it
is possible to let a coarse grained agent-based simulator interact
with a discrete event simulator such as~OMNeT++. It should be observed
that the same ideas can be applied in order to use other available
specific simulators, e.g., NS-3, Sumo, and so
forth~\cite{Brambilla:2014:SPL:2694768.2694780}.

%% file: sec_simulator.tex
\section{Multi-level Simulator for the IoT}\label{sec:simulator}

In this section, we describe the implementation of a multi-level
approach to simulate a general~\ac{IoT} scenario. The focus of this
paper is on a tool that allows simulating~\ac{IoT} applications,
rather than presenting a single application itself. Thus, in order to
show an instance of a multi-level simulator, we describe a general use
case with both static and mobile nodes immersed in a decentralized
area where no full cellular network coverage is available. Thus, other
communication approaches might be exploited in order to offer smart
services~\cite{smartshires,smartshires_abps}. Under some
circumstances, a subset of nodes might generate a MANET-like overlay;
in this case, from a simulation point of view it becomes mandatory
considering all the details related to wireless communications.

\subsection{Use Case}

We consider a scenario to foster the development of novel, smart
services in rural or decentralized
territories~\cite{smartshires,smartshires_abps}.  The aim is to
provide means to support proximity-based applications, that is,
applications where users can discover places of interest and ask to be
guided there from their current position. A conventional realization
of this service might be based on traditional client/server
approaches, where a user connects to given Web service, looks for some
point of interest, and then exploits a classic navigation tool to
reach it. However, scenarios exist where such a typical scheme cannot
be employed, because the user is located within an area with no full
infrastructured wireless connectivity coverage.  For instance, the
user might be located indoor or in a rural area lacking 3G/4G
connectivity. In this case, the use of an epidemic dissemination
protocol, or a MANET-like solution, might be of help.

As a use case for demonstrating the viability of employing multi-level
simulation for~\ac{IoT} environments, let us focus on a situation
where communications occur through ad-hoc networks only \cite{Ferretti2013481}. In order to
support the development of proximity-based applications, we need a
middleware service that allows disseminating messages looking for
users (located in the neighborhood) with specific interests.

A solution to build a data distribution service, in the absence of an
infrastructured network offering a complete coverage, might be that of
resorting to an unstructured, multi-hop relay based protocol (or
better, we might consider the idea of analyzing the performance of
such a kind of protocols, via multi-level simulation).  Hence, we have
a set of mobile and static nodes generating some content.  We assume
that these contents are treated as open data and are adequately
enriched with some kind of tags, keys or metadata, that allow nodes
filtering such data, upon reception.  Nodes are allowed to relay and
broadcast messages. On top of this dissemination strategy, it is
possible to implement a publish/subscribe approach that filters
information of interest for the applications.  Through this cheap and
simple dissemination strategy, producers are enabled to notify users
that subscribed to some specific keywords.

Once a subscriber has received some data that is of interest for him,
we can imagine that he moves towards a certain destination. During his
path, he is dynamically guided to the exact location through local
interactions with devices/sensors available in that area.

The simulation of such a wide scenario is not an easy task, since it
involves several activities and different domains. This is a perfect
example of a simulation scenario requiring different levels of
granularity. Thus, it is necessary to employ multi-level simulation.
This methodology is described in the next subsections.

\subsection{Multi-Level Simulator}

To simulate the use case presented above, we use a multi-level
approach that combines a discrete-event simulation engine (Level~1)
coupled with an agent-based model (Level~0).  The coarse level
(Level~0) simulates the whole~\ac{IoT} and related users, where
different actors produce data, subscribe their interests, move towards
different geographical areas. This has been implemented using an
agent-based simulator equipped with parallel and distributed execution
capabilities. The specific interactions within points of interest
(e.g., a marketplace) impose more simulation details to consider
wireless communication issues, fine-grained interactions and
movements.  Thus, a configurable OMNeT++ simulation model has been
implemented (Level 1). In this case, each simulation step of the
coarse grained simulation layer is decomposed into a set of events at
the fine grained layer. Following this approach, Level 1 is able to
notify Level 0 with its simulation advancements.

\subsection{Level 0: agent-based simulator}

The Smart Shire Simulator ($S^3$) is a prototype simulator based on
the GAIA/ART\`IS simulation
middleware~\cite{smartshires}. ART\`IS~\cite{artis} permits the
seamless sequential/parallel/distributed execution of large scale
simulation runs using different communication approaches (e.g.~shared
memory, TCP/IP, MPI) and synchronization methods (e.g.~time-stepped,
conservative, optimistic). The GAIA part~\cite{gda-simpat-2017} of the software
stack aims to ease the development of simulation models with high
level application program interfaces. Furthermore, it implements
communication and computational load-balancing strategies based on the
adaptive partitioning of the simulation model.

According to this simulator, entities in the~\ac{IoT} can be static
(e.g., sensors, traffic lights and road signs) or mobile entities
(e.g., cars and smartphones), following specific mobility models. All
these simulated things are equipped with a wireless interface
card. The interaction among entities is based on a ``Priority-based
Broadcast'' (PbB) strategy that implements a probabilistic broadcast
approach~\cite{1200529}. In~PbB, every message that is generated by a
node is broadcasted to all the nodes that are in proximity of the
sender. The message contains a Time-To-Live (TTL) to limit its
lifespan and the forwarding is based on two conditions. The first is a
probabilistic evaluation (i.e., probabilistic broadcast) while the
second is based on the distance between sender and receiver. In fact,
to limit the number of forwarded messages, there is a message forward
only if the distance between the nodes is larger than a given
threshold. Under the implementation viewpoint, this can be done using
a positioning system (e.g.,~GPS) if available. Otherwise, the network
signal level associated to each received message is used. Finally, a
message caching mechanism is employed to limit the presence of
duplicated messages.

\subsection{Level 1: OMNeT++ simulator}\label{sec:omnet}

The finer grained simulator has been implemented using OMNeT++
v.~4.4.1, with the INET framework v.~2.3.0. It simulates a grid of
fixed nodes (during the tests, a $10\times10$ grid was used),
representing a local subset of things/devices. Each device is equipped
with WiFi technology. In the simulated scenario, no WiFi
infrastructure was present, hence nodes organize themselves as a~MANET
exploiting DYMOUM~\cite{www_dymo-um}, an implementation of the Dynamic
MANET On-demand (DYMO) routing protocol~\cite{ietf-manet-dymo-26}.  

A number~$N$ of mobile nodes, representing pedestrian users, is
introduced by the higher Level~0 simulator. These~$N$ nodes are
equipped with a mobile device with WiFi connectivity.  Pedestrian
users move at walking speed. The user application running on the
mobile client broadcasts messages looking for the identifier of the
specific location. In fact, we imagine that the user destination
refers to a location (e.g., a point of interest, a seller within a
crowded farmers' market) where there is a device equipped with
communication capabilities, that is communicating in the~\ac{IoT}.
Thus, the device destination replies with his geographical
position. All these messages are delivered through the mentioned MANET
routing protocol. Based on the provided position, the mobile user
moves towards his destination.

\subsection{Interface between the two simulators}

A TCP-based message-passing approach is utilized to let the two
simulators communicate the inputs and outputs, as well as of
triggering the ``continue the simulation'' or ``end of simulation''
commands sent, at the end of each Level~0 timestep, from Level~0 to
Level~1.  In particular, at the end of each Level~0 timestep, Level~1
sends a set of messages which describe its status and waits for a
response. Level~0 receives the data sent by Level~1 and decides
whether Level~1 must continue or end the lower level simulation.

This is a simple strategy that enables interaction, and
synchronization, between the simulators without requiring a complete
re-engineering of the simulators. The higher levels simulator must be
able to freeze the simulation of certain parts of the scenario,
waiting for updates from other sources. Moreover, lower level
simulators should be enabled to obtain input from outside, and notify
results outside. However, no knowledge on the external simulators are
needed. This is an example demonstrating that existing products can be
employed to create more complex multilevel simulations.

%% file: sec_perf.tex
\section{Performance Evaluation}
\label{sec:perf}

The performance evaluation reported in this section refers to the
multi-level simulator as composed of the Level 0 and the Level 1
simulators described above.

The results are obtained averaging the outcomes of multiple
independent runs. The testbed used for the performance evaluation is
based on a DELL R620 with 2 CPUs (Xeon E-2640v2, 2 GHz, 8 physical
cores with Hyper-Threading). The software stack is composed of Ubuntu
14.04.5 LTS, GAIA/ART\`IS version 2.1.0, OMNeT++ v.~4.4.1 with the
INET framework v.~2.3.0.

As already mentioned, we implemented the simulation environment and
the application running on top of it. However, we are more interested
here on the effectiveness of the tool to simulate the considered
scenario, rather than the application itself. Put in other words, we
want to assess if this simulation strategy enables~\ac{IoT} developers
to effectively build simulation scenarios and test them in a
reasonable time, by considering large-scale~\ac{IoT} with high levels
of details, when needed.  For this reason, the main metrics of
interest is the Wall Clock Time (WCT), i.e., the time needed to
perform the simulation.

It is worth noting that, with the aim of comparing our solution with
another simulator offering the same level of detail in the simulation,
we built an OMNeT++ model comprising the whole simulation scenario
(not only the limited regions simulated where and when
necessary). Unfortunately, OMNeT++ was not able to scale up to the
amount of simulated entities shown in the following results. In fact,
simulations run with~$1000$ simulated entities lasted for
approximately 10 hours each. For this reason, in the next charts and
table we do not report results for that simulator.

\subsection{Level 0 simulator}

The Level~0 model is based on a bi-dimensional toroidal
space without obstacles. The simulated area is populated by a given
number of devices (in the following referred as Simulated Entities,
SEs). A part of the SEs is static while the others implement the well
known Random Way-Point (RWP)~\cite{rwp} mobility model. In the
simulated model, the communication is based on the proximity of SEs
and the multi-hop data dissemination protocol (previously described)
is implemented. Table~\ref{table:model} reports the main parameters of
the Level 0 simulation model.

Each SE uses a caching mechanism to discard some of the duplicated
messages generated by the PbB dissemination scheme. This caching
system is based on a LRU (Least Recently Used) replacement algorithm.

\begin{table}[h]
\begin{center}
  \begin{tabular}{ r p{4cm} }
    \toprule
    \textbf{Model parameter} & \textbf{Description/Value} \\ 
    \midrule
    Number of $SEs$ & $[1000, 32000]$ \\ 
    Mobility of $SEs$ & 50\% Random Way-point (RWP)\newline $50\%$ static  \\ 
    Speed of RWP & Uniform in the range $[1, 14]$ spaceunits/timestep \\ 
    Sleep time of RWP & $0$ (disabled) \\ 
    Interaction range & $250$ spaceunits \\ 
    Density of $SEs$ & $1$ node every $10000$ $\textrm{spaceunits}^2$ \\ 
    Forwarding range & $>200$ spaceunits \\ 
    Simulated time & $900$ timeunits \\ 
    Simulation granularity & $1$ timestep = $1$ timeunit \\ 
    Time-To-Live (TTL) & $4$ hops \\ 
    Dissemination probability & $0.6$ \\ 
    Cache size & $0$ (disabled) or $256$ \\ 
    \bottomrule
  \end{tabular}
\end{center}
\caption{Level~0 model parameters.}
\label{table:model}
\end{table}

\subsection{Level 1 simulator}

In this performance evaluation, the OMNeT++ simulation model described
in Section~\ref{sec:omnet} has been set up with a single SE transferred
from the Level~0 simulator to Level~1. The SE is managed by the
Level~1 for the length of a single Level~0 timestep. In other words,
at the beginning of the Level~0 timestep where a Level~1 simulation is
triggered, the Level~1 simulator is bootstrapped (including a warm-up
phase) and at the end of the Level~0 timestep the Level~1 simulator is
shutdown.

\subsection{Scalability evaluation}

\begin{figure*}[h]
\centering
\includegraphics[width=11.0cm]{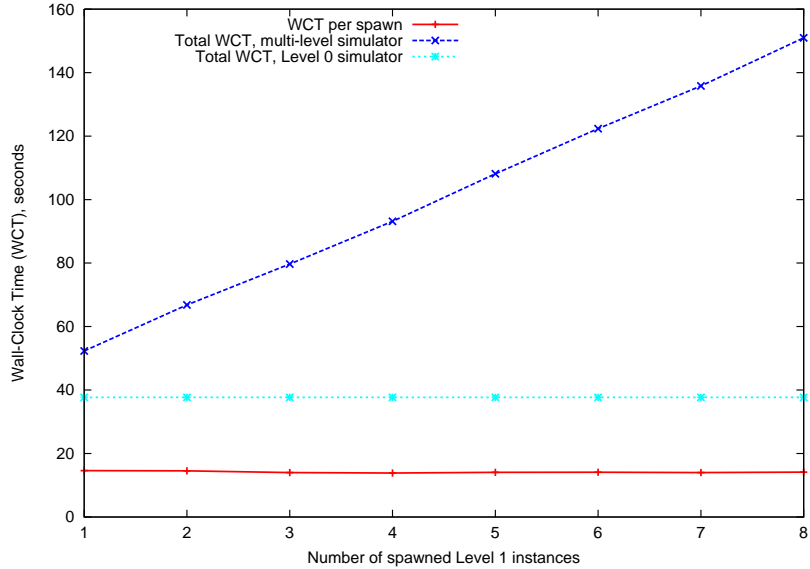}
\caption{WCT of the parallel multi-level simulator, sequential setup, 1000 SEs, increasing number of sequential spawned Level 1 instances.}
\label{fig:pads-lines-multiple_activations-SEQ}
\end{figure*}

First of all, we assess the scalability of the simulator in a
sequential setup, that is, $1$ CPU core is
used. Figure~\ref{fig:pads-lines-multiple_activations-SEQ} reports the
Wall-Clock Time (WCT) to complete a single simulation, on average. The
number of SEs has been set equal to~$1000$. In the graph, the
horizontal axis reports the amount of Level~1 instances that are
generated during a multi-level simulation; on the vertical axis we
show the~WCT. In particular, we report the average~WCT required to
complete the Level~0 simulation alone (cyan line), the average WCT
required by a single instance at Level~1 to complete its own (partial)
simulation (red line) and the average total~WCT needed to perform the
whole simulation (blue line). We observe that while each isolated
simulation requires a given WTC, the combination of Level~0 and
multiple Level~1 instances increases significantly the total~WCT.

The results show execution times of the order of few minutes, that
favorably compares to the several hours measured using a fine grained
OMNeT++ model alone, the tests described above show that sequential
simulations are unable to scale to the desired number of nodes and
level of detail. As a consequence, due to the scale of~\ac{IoT}
systems, the parallel setup of the multi-level simulator becomes of
main importance. Thus, the set of~SEs was partitioned among the CPU
cores and a message passing scheme was used to deliver the
interactions among~SEs. In this case, the software component executed
on each CPU core is called Logical Process (LP).

More in detail, we assume a configuration of the simulator with~$4$
LPs (run on~$4$ different CPU-cores) and in which, at given points in
the simulated time, all the LPs spawn a Level~1 instance. In other
words, multiple concurrent Level~1 instances are run at the same time.

\begin{figure*}[h]
\centering
\includegraphics[width=11.0cm]{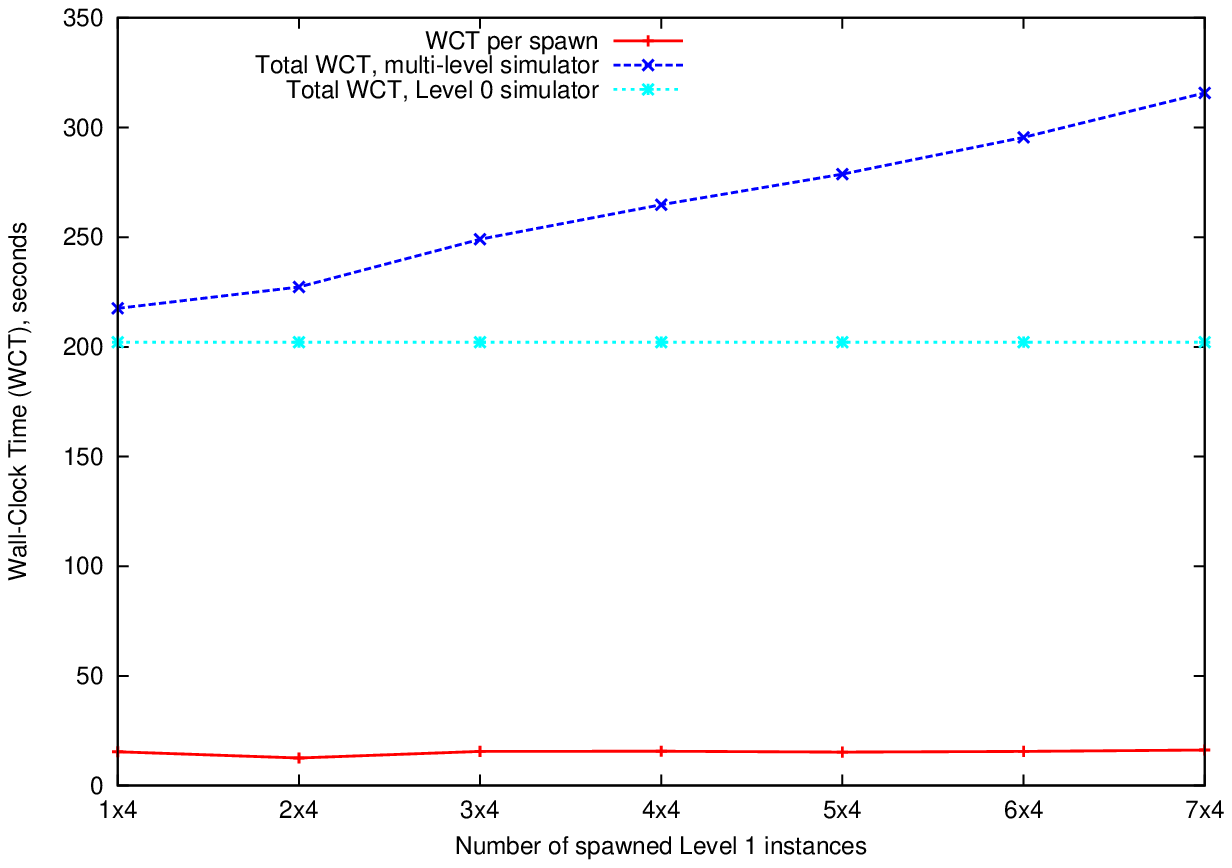}
\caption{WCT of the parallel multi-level simulator with $\#LPs=4$, 4000 SEs, increasing number of concurrent spawned Level~1 instances. For example, 2x4 means that all the LPs for 2 times during the simulation run trigger 4 concurrent Level 1 instances.}
\label{fig:pads-lines-multiple_activations}
\end{figure*}

Figure~\ref{fig:pads-lines-multiple_activations} shows the~WCT of the
multi-level simulator with an increasing number of concurrent
instances when~$4000$ SEs are simulated. As expected the activation of
Level~1 instances is costly but the scalability is still
reasonable. Furthermore, if in the parallel/distributed execution
architecture there are idle processor cores, then the execution of
concurrent Level~1 instances is quite cheap. In fact, they can run in
parallel with a negligible impact on the~WCT of the multi-level
simulator. Thus, these tests confirm that a good scalability can be
achieve through the proposed multi-level simulator.

Finally, we measured the amount of system memory used by the
multi-level simulator. Table~\ref{table:multi-memory-VmHWM} reports
the peak resident set size (i.e., VmHWM) used by the multi-level
simulator during the execution of a simulation run with an increasing
number of~SEs. The amount of~LPs was~$4$ in which, as before, at given
points in the simulated time, all the LPs spawn a Level~1 instance. It
is worth noting that, in this case, each Level~1 instance manages a
single~SE. Clearly, in this setup, the presence of four concurrent
Level~1 instances has a non-negligible impact on the total amount of
used memory, but the overall consumption is quite limited and in line
with our expectations. As a future work, both the Level~0 and Level~1
simulators will be optimized to reduce the memory consumption.  In any
case, also these measurements demonstrate that the use of multi-level
simulation allows simulating large scale~\ac{IoT} scenarios with high
levels of details in the simulation.

\begin {table}[th]
\begin{center}
    \begin{tabular}{lccc}
    \toprule
    $\#SEs$ & Level 0 (KB) & Level 1 (KB) per instance & Total (KB) \\ 
    \midrule
    1000  & 132880  & 43916	(x4) & 308544 \\ 
    2000  & 171728  & 43916	(x4) & 347392 \\ 
    4000  & 263488  & 43916	(x4) & 439152 \\ 
    8000  & 396416  & 43916	(x4) & 572080 \\ 
    16000 & 714720  & 43916	(x4) & 890384 \\ 
    32000 & 1351072 & 43916	(x4) & 1526736 \\ 
    \bottomrule
    \end{tabular}
    \caption{Peak resident set size used by the multi-level simulator;
      parallel setup with 4~LPs.}
    \label{table:multi-memory-VmHWM}
\end{center}
\end {table}

%% file: sec_conclusions.tex
\section{Conclusions}
\label{sec:conclusions}

In this paper we have presented a case study on multi-level simulation
of the~\acf{IoT}.  Our solution uses a two level simulation. An
adaptive, parallel/distributed agent-based simulation technique is
employed to model the coarse level (Level~0), while an OMNeT++
simulation implements the finer level (Level~1).

The performance results confirm the viability of the proposal. The use
of multi-level simulation allows scaling to a high numbers of SEs with
respect to the use of a single fine-grained simulator that is able to
capture the complexity and all the technical details of the
interactions among SEs. In fact, with a high number of~SEs, we
observed a total execution times of the simulation of the order of few
minutes, while the same simulations, built using a OMNeT++ model
alone, required several hours that became days when the number of
simulated entities increased.  The proposed approach allows mimicking
all the details only when needed. Hence, during the rest of the
simulation the multi-level simulator behaves as a coarse-grained
simulator.  Nevertheless, the interaction and synchronization among
active simulation levels guarantee a consistent and correct evolution
of the simulation.  To conclude, the presented approach represents a
viable tool to simulate specific IoT-based applications at a large
scale.

%% file: paper.bbl
\begin{thebibliography}{10}

\bibitem{iot_survey}
A.~Al-Fuqaha, M.~Guizani, M.~Mohammadi, M.~Aledhari, and M.~Ayyash.
\newblock Internet of things: A survey on enabling technologies, protocols, and
  applications.
\newblock {\em IEEE Communications Surveys Tutorials}, 17(4):2347--2376,
  Fourthquarter 2015.

\bibitem{Atzori:2010}
L.~Atzori, A.~Iera, and G.~Morabito.
\newblock The internet of things: A survey.
\newblock {\em Comput. Netw.}, 54(15):2787--2805, Oct. 2010.

\bibitem{bauza}
R.~Bauza, J.~Gozalvez, and M.~Sepulcre.
\newblock Operation and performance of vehicular ad-hoc routing protocols in
  realistic environments.
\newblock In {\em Vehicular Technology Conference, 2008. VTC 2008-Fall. IEEE
  68th}, pages 1--5, Sept 2008.

\bibitem{artis}
L.~Bononi, M.~Bracuto, G.~D'Angelo, and L.~Donatiello.
\newblock {\em ART{\`I}S: A Parallel and Distributed Simulation Middleware for
  Performance Evaluation}, pages 627--637.
\newblock Springer Berlin Heidelberg, Berlin, Heidelberg, 2004.

\bibitem{Brambilla:2014:SPL:2694768.2694780}
G.~Brambilla, M.~Picone, S.~Cirani, M.~Amoretti, and F.~Zanichelli.
\newblock A simulation platform for large-scale internet of things scenarios in
  urban environments.
\newblock In {\em Proc. of the First International Conference on IoT in Urban
  Space}, URB-IOT '14, pages 50--55, ICST, 2014.

\bibitem{Brumbulli2016}
M.~Brumbulli and E.~Gaudin.
\newblock {\em Proc. of the Second Asia-Pacific Conference on Complex Systems
  Design \& Management, CSD\&M Asia 2016}, chapter Towards Model-Driven
  Simulation of the Internet of Things, pages 17--29.
\newblock Springer, Cham, 2016.

\bibitem{gda-hpcs-16}
G.~D'Angelo, S.~Ferretti, and V.~Ghini.
\newblock Simulation of the internet of things.
\newblock In {\em 2016 International Conference on High Performance Computing
  Simulation (HPCS)}. IEEE, July 2016.

\bibitem{gda-hpcs-17}
G.~D'Angelo, S.~Ferretti, and V.~Ghini.
\newblock Modeling the internet of things: A simulation perspective.
\newblock In {\em Proc.~of the 2017 International Conference on High
  Performance Computing Simulation (HPCS)}. IEEE, July 2017.

\bibitem{gda-simpat-iot}
G.~D'Angelo, S.~Ferretti, and V.~Ghini.
\newblock Multi-level simulation of internet of things on smart territories.
\newblock {\em Simulation Modelling Practice and Theory (SIMPAT)}, 73:3 -- 21,
  2017.

\bibitem{gda-simpat-2014}
G.~D'Angelo and M.~Marzolla.
\newblock New trends in parallel and distributed simulation: From many-cores to
  cloud computing.
\newblock {\em Simulation Modelling Practice and Theory (SIMPAT)}, 2014.

\bibitem{www_dymo-um}
{DYMO-UM Project, http://masimum.dif.um.es/?Software:DYMOUM}.

\bibitem{gda-simpat-2017}
G.~D’Angelo.
\newblock The simulation model partitioning problem: an adaptive solution based
  on self-clustering.
\newblock {\em Simulation Modelling Practice and Theory (SIMPAT)}, 70:1 -- 20,
  2017.

\bibitem{Ferretti2013481}
S.~Ferretti.
\newblock Shaping opportunistic networks.
\newblock {\em Computer Communications}, 36(5):481 -- 503, 2013.

\bibitem{smartshires}
S.~Ferretti and G.~D'Angelo.
\newblock Smart shires: The revenge of countrysides.
\newblock In {\em Proceedings of the IEEE Symposium on Computers and
  Communications}, ISCC '16, Washington, DC, USA, 2016. IEEE Computer Society.

\bibitem{smartshires_abps}
S.~Ferretti, G.~D'Angelo, and V.~Ghini.
\newblock Smart multihoming in smart shires: Mobility and communication
  management for smart services in countrysides.
\newblock In {\em Proceedings of the IEEE Symposium on Computers and
  Communications}, ISCC '16, Washington, DC, USA, 2016. IEEE Computer Society.

\bibitem{6069710}
A.~Gluhak, S.~Krco, M.~Nati, D.~Pfisterer, N.~Mitton, and T.~Razafindralambo.
\newblock A survey on facilities for experimental internet of things research.
\newblock {\em IEEE Communications Magazine}, 49(11):58--67, November 2011.

\bibitem{6803226}
S.~N. Han, G.~M. Lee, N.~Crespi, K.~Heo, N.~V. Luong, M.~Brut, and
  P.~Gatellier.
\newblock Dpwsim: A simulation toolkit for iot applications using devices
  profile for web services.
\newblock In {\em Internet of Things (WF-IoT), 2014 IEEE World Forum on}, pages
  544--547, March 2014.

\bibitem{Karnouskos}
S.~Karnouskos and T.~N. d.~Holanda.
\newblock Simulation of a smart grid city with software agents.
\newblock In {\em Computer Modeling and Simulation, 2009. EMS '09. Third UKSim
  European Symposium on}, pages 424--429, Nov 2009.

\bibitem{kerekes}
J.~P. Kerekes, M.~D. Presnar, K.~D. Fourspring, Z.~Ninkov, D.~R. Pogorzala,
  A.~D. Raisanen, A.~C. Rice, J.~R. Vasquez, J.~P. Patel, R.~T. MacIntyre, and
  S.~D. Brown.
\newblock Sensor modeling and demonstration of a multi-object spectrometer for
  performance-driven sensing.
\newblock volume 7334, pages 73340J--73340J--12, 2009.

\bibitem{paper+kirsche-13:iot-simulation}
M.~Kirsche.
\newblock Simulating the internet of things in a hybrid way.
\newblock In {\em Proceedings of the Networked Systems (NetSys) 2013 PhD
  Forum}, 03 2013.
\newblock Poster Abstract.

\bibitem{SUMO2012}
D.~Krajzewicz, J.~Erdmann, M.~Behrisch, and L.~Bieker.
\newblock Recent development and applications of {SUMO - Simulation of Urban
  MObility}.
\newblock {\em International Journal On Advances in Systems and Measurements},
  5(3\&4):128--138, December 2012.

\bibitem{6664581}
V.~Looga, Z.~Ou, Y.~Deng, and A.~Ylä-Jääski.
\newblock Mammoth: A massive-scale emulation platform for internet of things.
\newblock In {\em 2012 IEEE 2nd International Conference on Cloud Computing and
  Intelligence Systems}, volume~03, pages 1235--1239, Oct 2012.

\bibitem{Luke:2005}
S.~Luke, C.~Cioffi-Revilla, L.~Panait, K.~Sullivan, and G.~Balan.
\newblock Mason: A multiagent simulation environment.
\newblock {\em Simulation}, 81(7):517--527, July 2005.

\bibitem{magne2000towards}
L.~Magne, S.~Rabut, and J.-F. Gabard.
\newblock Towards an hybrid macro-micro traffic flow simulation model.
\newblock In {\em INFORMS spring 2000 meeting}, 2000.

\bibitem{rwp}
M.~Musolesi and C.~Mascolo.
\newblock Mobility models for systems evaluation.
\newblock In {\em Middleware for Network Eccentric and Mobile Applications},
  pages 43--62. Springer Berlin Heidelberg, 2009.

\bibitem{ietf-manet-dymo-26}
C.~E. Perkins, S.~Ratliff, and J.~Dowdell.
\newblock {Dynamic MANET On-demand (AODVv2) Routing}.
\newblock Internet-Draft draft-ietf-manet-dymo-26, Internet Engineering Task
  Force, Apr. 2013.
\newblock Work in Progress.

\bibitem{1200529}
Y.~Sasson, D.~Cavin, and A.~Schiper.
\newblock Probabilistic broadcast for flooding in wireless mobile ad hoc
  networks.
\newblock In {\em Wireless Communications and Networking}, volume~2, pages
  1124--1130 vol.2, March 2003.

\bibitem{singh2014survey}
D.~Singh, G.~Tripathi, and A.~J. Jara.
\newblock A survey of internet-of-things: Future vision, architecture,
  challenges and services.
\newblock In {\em Internet of things (WF-IoT), 2014 IEEE world forum on}, pages
  287--292. IEEE, 2014.

\bibitem{6418824}
Y.~Song, B.~Han, X.~Zhang, and D.~Yang.
\newblock Modeling and simulation of smart home scenarios based on internet of
  things.
\newblock In {\em IEEE Int. Conf. on Network Infrastructure and Digital
  Content}, pages 596--600, Sept 2012.

\bibitem{6844677}
S.~Sotiriadis, N.~Bessis, E.~Asimakopoulou, and N.~Mustafee.
\newblock Towards simulating the internet of things.
\newblock In {\em 28th Int. Conf. on Advanced Information Networking and
  Applications Workshops (WAINA)}, pages 444--448, May 2014.

\bibitem{Wegener:2008}
A.~Wegener, M.~Pi\'{o}rkowski, M.~Raya, H.~Hellbr\"{u}ck, S.~Fischer, and J.-P.
  Hubaux.
\newblock Traci: An interface for coupling road traffic and network simulators.
\newblock In {\em Proc. of the 11th Communications and Networking Simulation
  Symposium}, CNS '08, pages 155--163. ACM, 2008.

\bibitem{e16052384}
D.~Zubillaga, G.~Cruz, L.~D. Aguilar, J.~Zapotécatl, N.~Fernández,
  J.~Aguilar, D.~A. Rosenblueth, and C.~Gershenson.
\newblock Measuring the complexity of self-organizing traffic lights.
\newblock {\em Entropy}, 16(5):2384, 2014.

\end{thebibliography}
